\documentclass[twocolumn,prl,amsmath,amssymb,showpacs,superscriptaddress,floatfix]{revtex4}
\usepackage{graphicx}
\usepackage{bm}

\usepackage{epsf}
\begin{document}
\title{Influence of random pinning on melting scenario of two-dimensional core-softened potential system}

\author{E. N. Tsiok, Yu. D. Fomin, and V. N. Ryzhov}
\affiliation{ Institute for High Pressure Physics RAS, 108840
Kaluzhskoe shosse, 14, Troitsk, Moscow, Russia}




\date{\today}

\begin{abstract}
The random disorder can drastically change the melting scenario of
two-dimensional systems and has to be taken into account in the
interpretation of the experimental results. We present the results
of the molecular dynamics simulations of the two dimensional
system with the core-softened potentials with two repulsive steps
in which a small fraction of the particles is pinned, inducing
quenched disorder. It is shown that without the quenched disorder
the system with small repulsive shoulder, which is close in the
shape to the soft disks, melts in accordance with the melting
scenario proposed in Refs.~\cite{foh1,foh2,foh4} (first-order
liquid-hexatic and continuous hexatic-solid transitions). Random
pinning widens the hexatic phase, but leaves the melting scenario
unchanged. For the system with larger repulsive step at high
densities the conventional first-order transition takes place
without random pinning. However, in the presence of disorder the
single first-order transition transforms into two transitions, one
of them (solid-hexatic) is the continuous KTHNY-like transition,
while the hexatic to isotropic liquid transition occurs as the
first order transition.
\end{abstract}


\maketitle

\section{Introduction}

Nature of two-dimensional melting is one of the long-standing
problems in condensed matter physics. Despite a lot of
publications in past forty years, the controversy about the
microscopic nature of melting in two dimensions ($2D$) still
lasts. In contrast to the three dimensional case, where melting
occurs through the conventional first order transition, several
microscopic scenarios were proposed for the microscopic
description of $2D$ melting. The reason of this difference is the
drastic increase of fluctuations in $2D$ in comparison with the
$3D$ case. Landau, Peierls and Mermin \cite{mermin} showed that in
two dimensions the long-ranged positional order can not exist
because of the thermal fluctuations and transforms to the
quasi-long-ranged one. On the other hand, the real long-ranged
orientational order (the order in directions of the bonds joining
the particle with its nearest neighbors) does exist in this case.

Now the Kosterlitz-Thouless-Halperin-Nelson-Young (KTHNY) theory
of $2D$ melting \cite{kosthoul73,halpnel1,halpnel2,halpnel3} is
the most widely accepted. In this theory it was proposed that
$2D$ solids melt through two continuous transitions, which are
induced by the formation of the topological defects. For example,
an elementary topological defect of the triangular lattice is a
disclination. Disclination in a triangular crystal lattice is
defined as an isolated defect with five or seven nearest
neighbours. A dislocation can be considered as a bound pair of 5
and 7-fold disclinations. In the framework of the KTHNY scenario,
the $2D$ solid melts through dissociation of bound dislocation
pairs. In this case the long-ranged orientational order transforms
into quasi-long-ranged one, and quasi-long-ranged positional order
becomes short-ranged. This new intermediate phase is called a
hexatic phase. In turn, the hexatic phase transforms into an
isotropic liquid phase with a short-ranged orientational and
positional orders through unbinding of dislocations (5 and 7-fold
bound pairs) into free disclinations. It should be noted, that the
KTHNY theory gives only limits of stability of the solid and
hexatic phases and does not reject the possibility of the
preempting first order liquid-solid transition with another
melting mechanism.

The KTHNY theory seems universal and applicable to all system.
However, it contains two phenomenological parameters, the core
energy of dislocation $E_c$ and Frank module of hexatic phase
$K_A$, which do not have explicit expressions in terms of the
interparticle potential. As it was shown later, with decreasing
$E_c$ melting can occur through a single first-order transition as
a result of, for example, the formation of grain boundaries
\cite{chui83} or "explosion" of 5-7-5-7 quartets (bound
dislocation pairs) into free disclinations \cite{ryzhovJETP}.

Recently the KTNHY scenario was unambiguously experimentally
confirmed for superparamagnetic colloidal particles interacting
via long-range dipolar interaction
\cite{keim1,zanh,keim2,keim3,keim4}. In these experiments the
particles are absorbed at liquid-air interface, which restricts
the out-of-plane motion. On the other hand, in very popular
experimental systems of colloidal particles confined between two
glass plates \cite{col1} the melting transition consistent with
the KTHNY scenario was found \cite{col2,col3,col4}. At the same
time, a first order liquid-solid  \cite{col5} and even a
first-order liquid-hexatic and a first-order hexatic-solid phase
transitions \cite{col6}  are also possible. One can conclude from
these results that the melting mechanism is not universal and
depends on the interparticle interactions. However, controversies
still exist even for the same systems, like, for example, for hard
spheres
\cite{rto1,rto2,RT1,RT2,DF,strandburg92,binderPRB,mak,binder,LJ}.

Recently, another melting scenario was proposed
\cite{foh1,foh2,foh3,foh4,foh5,foh6}. In contrast to the KTHNY
theory, it was argued that in the basic hard disk model the
hexatic phase does exists, and the system melts through  a
continuous solid-hexatic transition but a first-order
hexatic-liquid transition \cite{foh1,foh2,foh3}. In Ref.
\cite{foh5} it was shown that there is a first-order transition
between the stable hexatic phase and isotropic liquid in $2D$
Yukawa system. In their work \cite{foh4} Kapfer and Krauth have
considered the behavior of soft disk system with the potential of
the form $U(r)=(\sigma/r)^n$. They have shown that the system
melts in accordance with the KTHNY theory for $n\leq 6$, while for
$n>6$ the two-stage melting transition takes place with the
continuous solid-hexatic transition and the first-order
hexatic-liquid one. Experimental confirmation of this scenario was
found in the system of colloidal particles on water-decane
interface \cite{col7}. It should be also noted the recent
publication \cite{foh7}, where for the Herzian disks model it was
shown that at low densities there is a discontinuous
liquid-hexatic transition while at higher densities the system
melts through the continuous Kosterlitz-Thouless transition.

In real experiments, two-dimensional confinement is typically
realized in slit pores of different nature or by adsorption on
solid substrates. In both cases the frozen-in (quenched) disorder
due to some roughness or intrinsic defects can appear. Quenched
disorder can change the melting scenario in $2D$. A disordered
substrate  can have the similar destructive effect on the
crystalline order as temperature and can lead to the melting even
at zero temperature \cite{nel_dis1,nel_dis2,dis3,dis4}. In Refs.
\cite{nel_dis1,nel_dis2} it was shown, that the KTHNY melting
scenario persists in the presence of weak disorder. As it is
intuitively clear, the temperature of the hexatic-isotropic liquid
transition $T_i$ is almost unaffected by disorder, while the
melting temperature $T_m$ drastically decreases with increasing
disorder \cite{dis3,dis4,nel_dis1,nel_dis2}. As a result, the
stability range of the hexatic phase widens. These predictions
have been confirmed in experiment and simulations of the system of
superparamagnetic colloidal particles \cite{keim3,keim4}. In these
experiments the particles form a monolayer on the bottom of a
cylindrical glass cell of $5-mm$ diameter due to gravity. Quenched
disorder occurs due to pinning of a small amount of particles to
the glass substrate due to van der Waals interactions and chemical
reactions. In our simulations we tried to choose the simulation
method in qualitative accordance with this experimental setup.

In this paper, we present a detailed computer simulation study of
$2D$ phase diagram of the previously introduced core softened
potential system \cite{jcp2008,wepre,we_inv,we2011,RCR,we2013-2}
in the presence of the quenched disorder for different values of
the width of the repulsive shoulder. Various forms of the
core-softened potentials are widely used for the qualitative
description of the water-like anomalous behavior, including
density, structural and diffusion anomalies, liquid-liquid phase
transitions, glass transitions, melting maxima
\cite{jcp2008,wepre,we_inv,we2011,RCR,we2013-2,buld2009,fr1,fr2,fr3,fr4,
barbosa,barbosa1,barbosa2,barbosa3,buld2d,scala,prest2,prest1}. In
our previous publications the preliminary result on the phase
diagram of the system were reported
\cite{dfrt1,dfrt2,dfrt3,dfrt4}. It was shown that
the random pinning widens the range of the hexatic phase  and
transforms the first-order melting at high densities into two
transitions - first-order liquid-hexatic and continuous
Kosterlitz-Thouless-like solid-hexatic transitions \cite{dfrt5}.
Here we present the corrected version of the phase diagram for the
system with the larger repulsive shoulder, obtained with the use
of additional criteria including the calculation of the diffusion
coefficient of the hexatic phase.

We also study the system with small repulsive shoulder. It is
shown, that in this case the behavior of the system is similar to
the soft disks one \cite{foh4}. The presence of the random pinning
leads to drastic increase of the width of the hexatic phase, but
does not change the melting scenario.

\section{Systems and methods}

\begin{figure}
\begin{center}
\includegraphics[width=8cm]{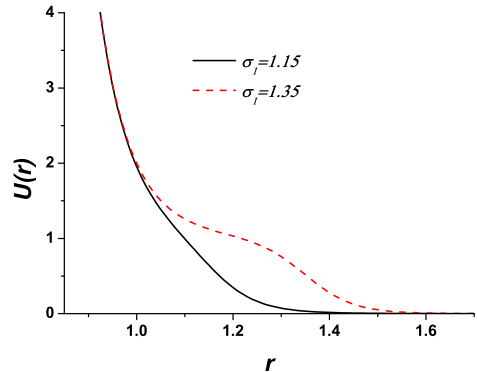}%

\end{center}

\caption{\label{fig:fig1} The potential (\ref{3}) with two
different soft-core diameters: $\sigma_1/\sigma = 1.15; 1.35$.}
\end{figure}

In the present simulations we study the system which is described
by the potential
\cite{jcp2008,wepre,we_inv,we2011,RCR,we2013-2,dfrt1,dfrt2,dfrt3,dfrt5}:
\begin{equation}
U(r)=\varepsilon\left(\frac{\sigma}{r}\right)^{n}+\frac{1}{2}\varepsilon\left(1-
\tanh(k_1\{r-\sigma_1\})\right). \label{3}
\end{equation}
where $n = 14$ and $k_1\sigma = 10.0$. $\sigma$ is the hard-core
diameter. We simulate the systems with two different soft-core
diameters: $\sigma_1/\sigma = 1.15; 1.35$. (see
Fig.~\ref{fig:fig1}).

In the remainder of this paper we use the dimensionless
quantities, which in $2D$ have the form: $\tilde{{\bf r}}\equiv
{\bf r}/\sigma$, $\tilde{P}\equiv P \sigma^{2}/\varepsilon ,$
$\tilde{V}\equiv V/N \sigma^{2}\equiv 1/\tilde{\rho}, \tilde{T}
\equiv k_BT/\varepsilon, \tilde{\sigma}=\sigma_1/\sigma$. In the
rest of the article the tildes will be omitted.

In this article we make the molecular dynamics simulations  in
$NVT$ and $NVE$ ensembles in the framework of the LAMMPS package
\cite{lammps} with the number of particles from $20000$ to
$100000$. In order to obtain the quenched disorder in the system,
we randomly choose a subset of particles at the random positions
and let them to be immobile for the entire simulation run
\cite{dfrt5}. The simulations of $10$ independent replicas of the
system with different distributions of random pinned patterns were
made. The thermodynamic functions were calculated by averaging
over replicas. We calculate the pressure $P$ versus density $\rho$
along the isotherms, and the correlation functions $G_6(r)$ and
$G_T(r)$ of the bond orientational $\psi_6$ and translational
$\psi_T$ order parameters (OPs), which characterize the overall
orientational and translational order \cite{dfrt5}.

The translational $\psi_T$ (TOP) and the orientational order
parameters $\psi_6$ (OOP) along with the bond-orientational
$G_6(r)$ (OCF) and translational $G_T(r)$ (TCF) correlation
functions are defined in the conventional way
\cite{prest2,halpnel1,halpnel2,binder,binderPRB,LJ,prest1,foh7}
with the subsequent averaging over the quenched disorder
\cite{dfrt5}.

In accordance with the standard definitions \cite{halpnel1,
halpnel2, foh7}, TOP has the form:
\begin{equation}
\psi_T=\frac{1}{N}\left<\left<\left|\sum_i e^{i{\bf G
r}_i}\right|\right>\right>_{rp}, \label{psit}
\end{equation}
where ${\bf r}_i$ is the position vector of particle $i$ and {\bf
G} is the reciprocal-lattice vector of the first shell of the
crystal lattice. The translational correlation function can be
obtained from the equation:
\begin{equation}
G_T(r)=\left<\frac{<\exp(i{\bf G}({\bf r}_i-{\bf
r}_j))>}{g(r)}\right>_{rp}, \label{GT}
\end{equation}
where $r=|{\bf r}_i-{\bf r}_j|$ and $g(r)=<\delta({\bf
r}_i)\delta({\bf r}_j)>$  is the pair distribution function. The
second angular brackets $<...>_{rp}$ correspond to the averaging
over the random pinning. In the solid phase without random pinning
the long range behavior of $G_T(r)$ has the form $G_T(r)\propto
r^{-\eta_T}$ with $\eta_T \leq \frac{1}{3}$ \cite{halpnel1,
halpnel2}.

To measure the orientational order and the hexatic phase, the
local order parameter which determines the $6$-fold orientational
ordering can be defined in the following way:
\begin{equation}
\Psi_6({\bf r_i})=\frac{1}{n(i)}\sum_{j=1}^{n(i)} e^{i
n\theta_{ij}}\label{psi6loc},
\end{equation}
where $\theta_{ij}$ is the angle of the vector between particles
$i$ and $j$ with respect to a reference axis and the sum over $j$
is counting $n(i)$ nearest-neighbors of $j$, obtained from the
Voronoi construction. The global OOP can be calculated as an
average over all particles and random pinning:
\begin{equation}
\psi_6=\frac{1}{N}\left<\left<\left|\sum_i \Psi_6({\bf
r}_i)\right|\right>\right>_{rp}.\label{psi6}
\end{equation}

The orientational correlation function $G_6(r)$ (OCF) is given in
the similar way as  in Eq. (\ref{GT}):
\begin{equation}
G_6(r)=\left<\frac{\left<\Psi_6({\bf r})\Psi_6^*({\bf
0})\right>}{g(r)}\right>_{rp}, \label{g6}
\end{equation}
where $\Psi_6({\bf r})$ is the local bond-orientational order
parameter (\ref{psi6loc}). In the hexatic phase there is a
quasi-long range order with the algebraic decay of the
orientational correlation function $G_6(r) \propto r^{-\eta_6}$
with $0\leq \eta_6 \leq \frac{1}{4}$
\cite{halpnel1,halpnel2,halpnel3}. The stability criterion of the
hexatic phase has the form $\eta_6(T_i) = \frac{1}{4}$.

The influence of disorder on the orientational and translational
correlation functions has been considered in our previous
publication \cite{dfrt5} (see Fig. 1 in \cite{dfrt5}). In the
presence of pinning, it was not found any qualitative change in
the behavior of $G_6(r)$ in accordance with the results of Nelson
and coworkers \cite{nel_dis1, nel_dis2} and our intuitive
expectations. On the other hand, the translational correlation
function $G_T(r)$ demonstrates the qualitatively different
behavior when the random disorder presents. Without pinning one
has the conventional power law decay of TCF. In the case of
pinning, the slope of the enveloping line increases at some
crossover value $r_0$. The region $r<r_0$ corresponds to the local
order unaffected by quenched disorder, while asymptotic behavior
of TCF for $r>r_0$ is determined by the random pinning
\cite{dfrt5}. It was also shown that, in accordance with the
intuitive physical picture, $r_0$ decreases with increase of
impurities concentration along with the increase of the slope of
the enveloping line. The equality $\eta_T(T_m)=1/3$ for the long
range asymptote of TCF (for $r>r_0$) can be considered as the
solid-hexatic stability criterion. The hexatic-liquid stability
point corresponds to $\eta_6(T_i)=1/4$ \cite{nel_dis1,nel_dis2}.

\begin{figure}

\includegraphics[width=8cm]{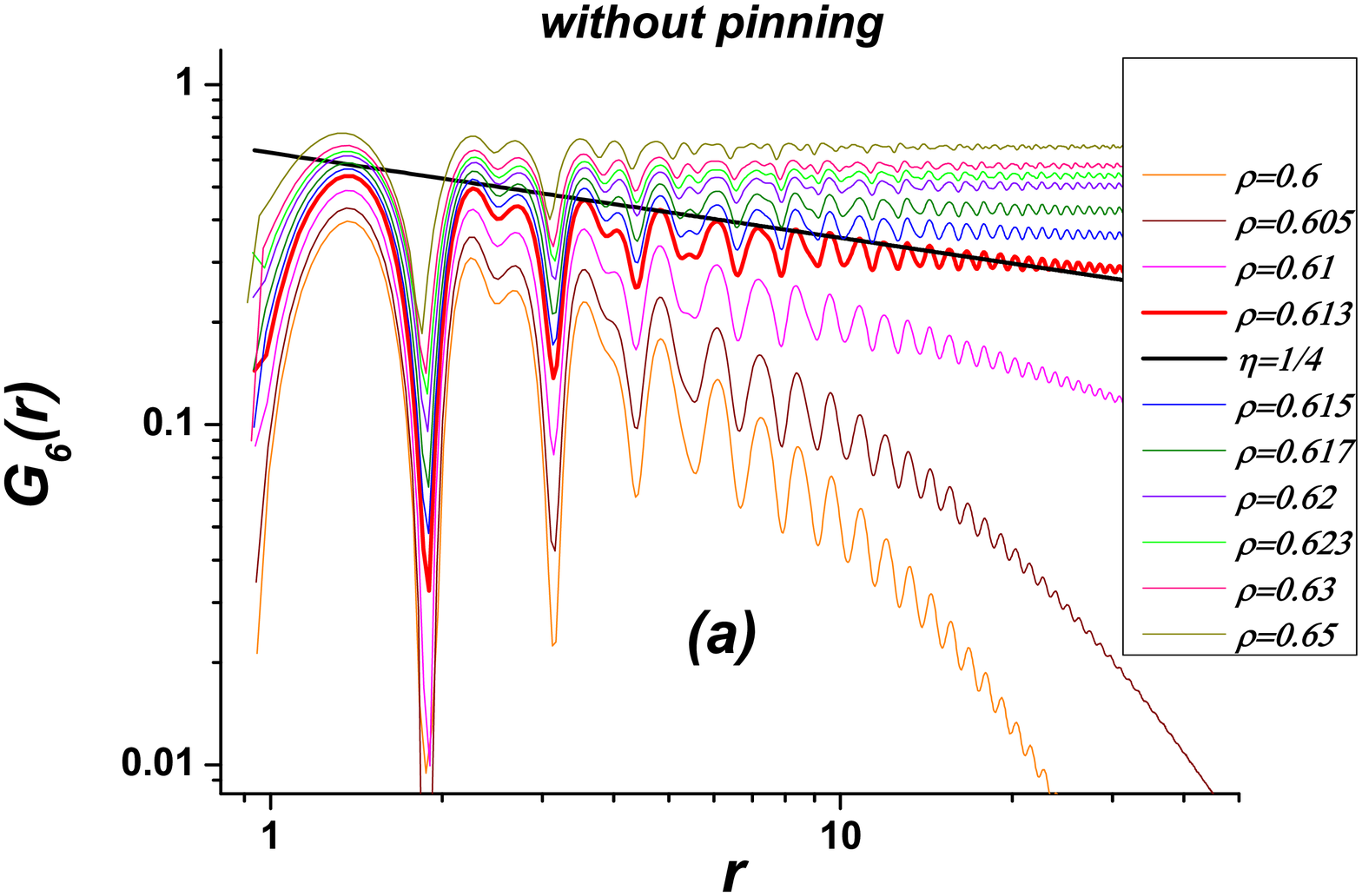}%

\includegraphics[width=8cm]{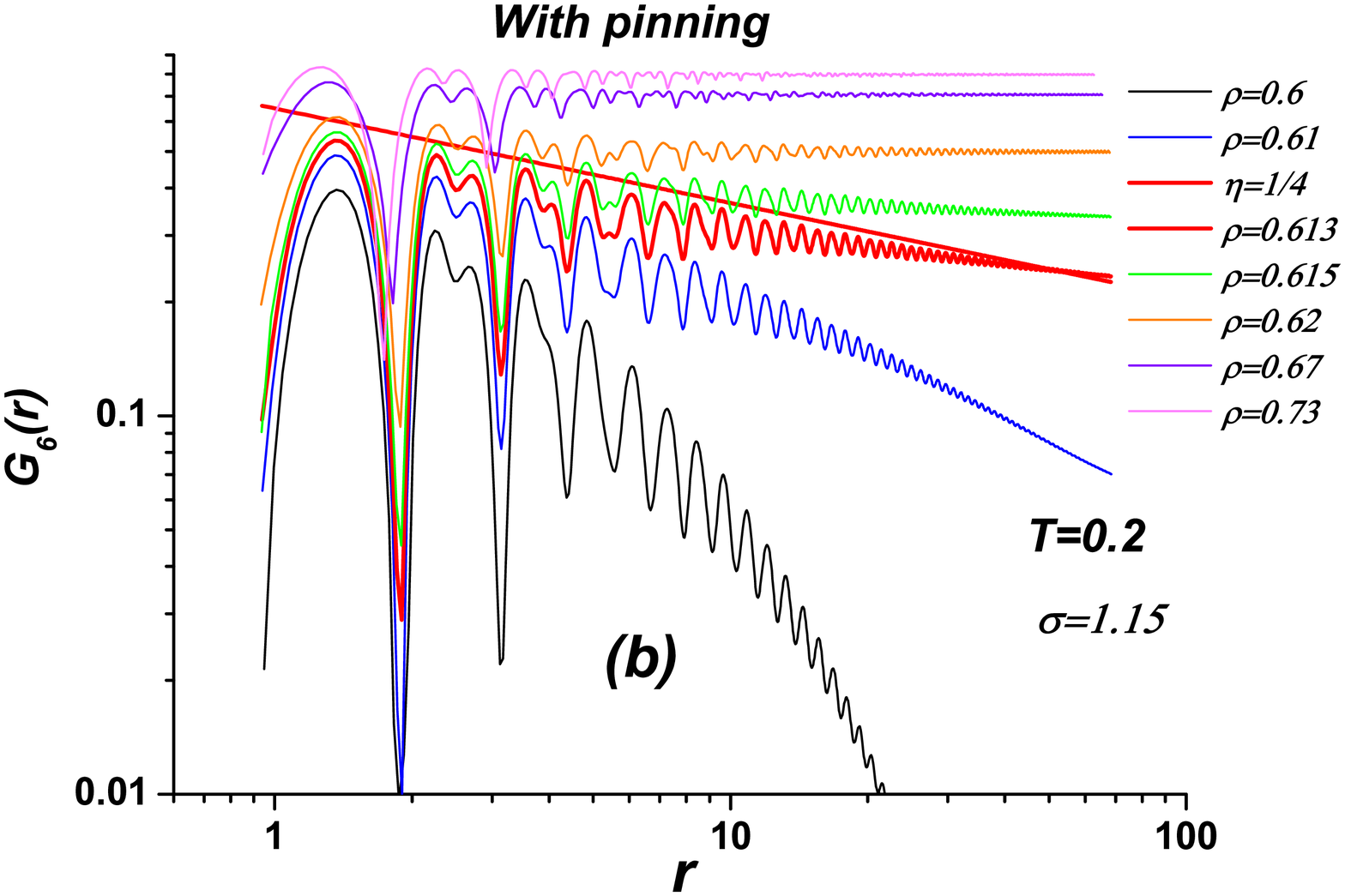}%

\caption{\label{fig:fig2} OCF  for different densities at $T=0.2$
without random pinning (a) and concentration of the pinning
centers $0.1\%$ (b).}
\end{figure}

\section{results and discussion}
\subsection{$\sigma=1.15$}

Let us first consider the behavior of the system with the small
repulsive shoulder $\sigma=1.15$. In this case, as one can see in
Fig.~\ref{fig:fig1}, the form of the potential is very close to
the soft disk with $n=14$. The preliminary discussion of the phase
diagram of this system can be found in Ref. \cite{dfrt2}, where
the isotherms with the  van der Waals loops are presented along
with the phase diagram calculated with the help of the
double-tangent construction to the Helmholtz free energy curves
\cite{book_fs}.

In Fig.~\ref{fig:fig2} we show the orientational correlation
functions (OCF) for the system without quenched disorder (a) and
with the random pinning (concentration of pinning centers is
$0.1\%$) at temperature $T=0.2$. One can see that the behavior of
the OCFs is equivalent for both cases, as it was discussed in the
Introduction, and the limits of stability of the hexatic phase,
determined from the condition $\eta=1/3$, coincide too. From
Fig.~\ref{fig:fig2} it follows that the density at which hexatic
phase becomes unstable is $\rho\approx 0.613$.

\begin{figure}%

\includegraphics[width=8cm]{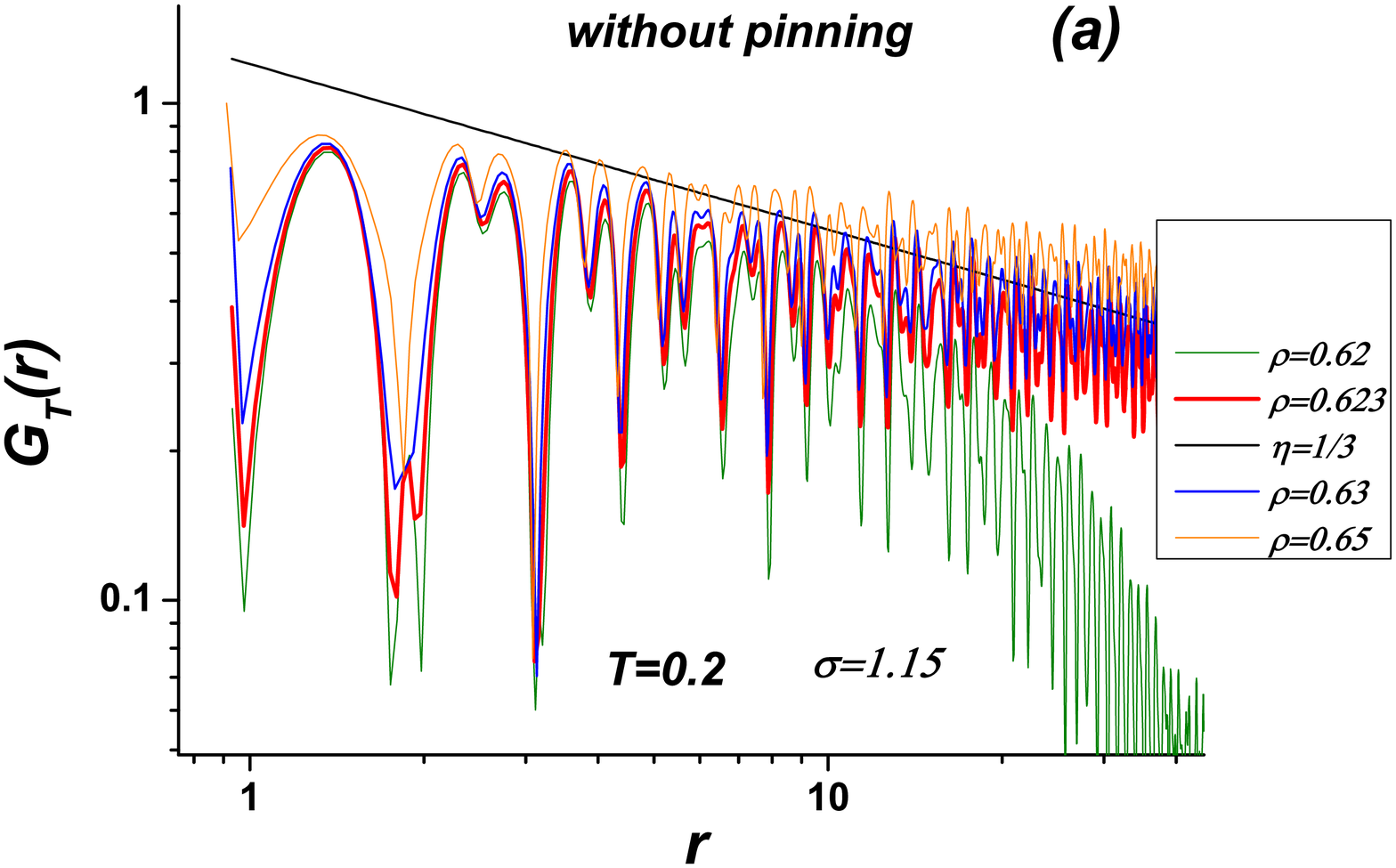}%

\includegraphics[width=8cm]{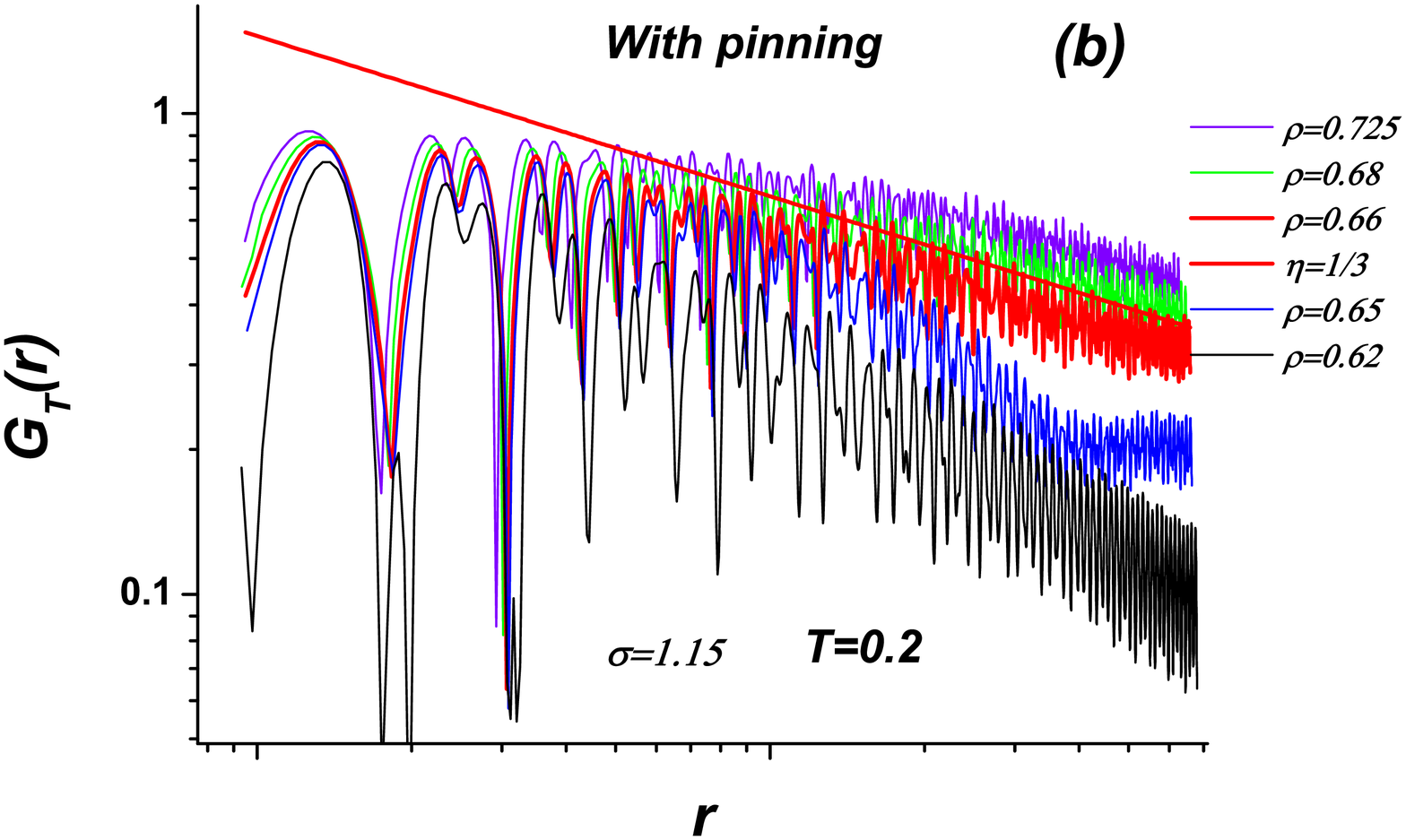}%

\caption{\label{fig:fig3}  TCF  for different densities at $T=0.2$
without random pinning (a) and concentration of the pinning
centers $0.1\%$ (b).}
\end{figure}

\begin{figure}%

\includegraphics[width=8cm]{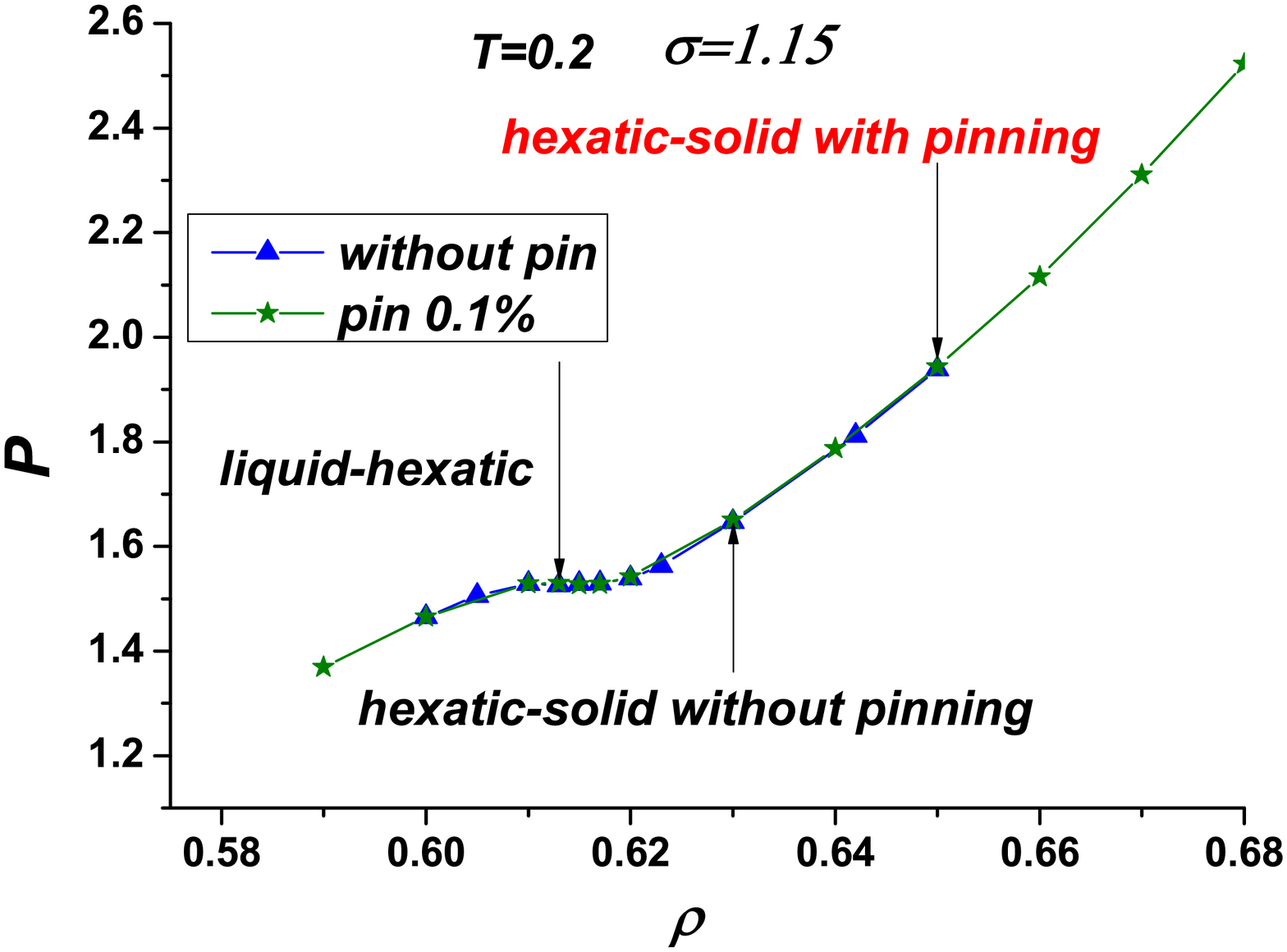}%

\caption{\label{fig:fig4} Isotherms of the system with
$\sigma=1.15$ without pinning (trianges) and with the
concentration of the pinning centers equal to $0.1\%$ (stars) for
$T=0.2$.}
\end{figure}

\begin{figure}
\begin{center}
\includegraphics[width=8cm]{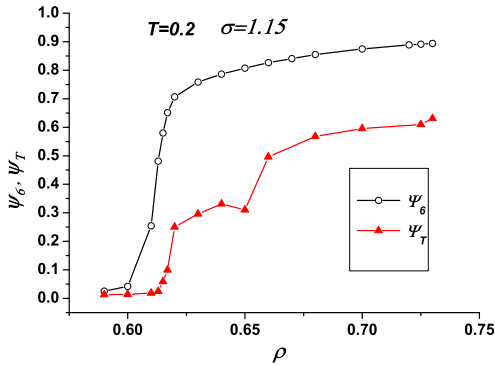}%

\end{center}

\caption{\label{fig:fig5}  The orientational $\Psi_6$ and
translational $\Psi_T$ order parameters for the system with random
pinning and $\sigma = 1.15$.}
\end{figure}

\begin{figure}
\begin{center}
\includegraphics[width=8cm]{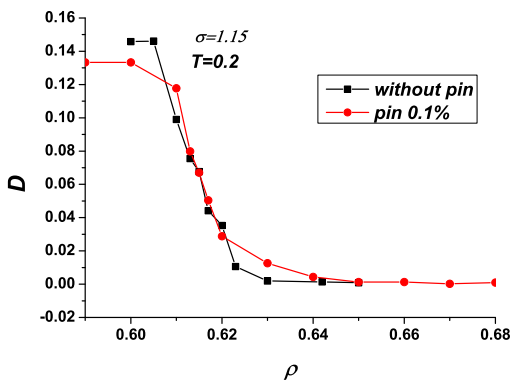}%

\end{center}

\caption{\label{fig:fig6}  The diffusion coefficient of the system
with $\sigma = 1.15$ with (red circles) and without (black
squares) random pinning.}
\end{figure}

On the other hand, the translational correlation functions of the
systems with and without random pinning are qualitatively
different (see Fig.~\ref{fig:fig3}). In this case the densities
corresponding to the loss of stability of the solid phase are
considerably different. From Fig.~\ref{fig:fig3} (a) one can see
that without pinning the stability limit corresponds to
$\rho\approx 0.63$ while in the system with the quenched disorder
the solid-hexatic stability limit is $\rho\approx 0.65$ (see
Fig.~\ref{fig:fig3} (b)).

In Fig.~\ref{fig:fig4} the equation of state at $T=0.2$ is
presented. It is seen that the liquid-hexatic stability limit
density is located inside the Van der Waals loop region for both
systems - with and without the random pinning. At the same time,
for the pinning-free system the hexatic-solid stability limit is
located outside the Van der Waals loop. One can conclude that in
this case the system melts in accordance with the melting scenario
proposed in Refs.~\cite{foh1,foh2,foh5}.

In the presence of the random pinning the hexatic-solid
instability density shifts to the higher densities, however, the
melting scenario does not change. We have first-order
liquid-hexatic transition and continuous hexatic-solid transition.
In Fig.~\ref{fig:fig5} the corresponding behavior of the
orientational $\psi_6$ and translational $\psi_T$ order parameters
is shown. The orientational order parameter $\psi_6$ demonstrates
the conventional behavior characteristic, for example, for the
system without random pinning \cite{dfrt2}. However, the
translational order parameter $\psi_T$ has the qualitatively
different form. In Fig.~\ref{fig:fig5} one can see that the
$\psi_T$ curve has step-like behavior. In decreasing the density,
$\psi_T$ jumps downward at $\rho\approx 0.65$ corresponding to the
stability limit of the solid phase in the presence of random
pinning. On the other hand, $\psi_T$ is not equal to zero at
$\rho\approx 0.65$ because the local translational order does
exists in this state (see Fig.~\ref{fig:fig3} and discussion in
the previous Section). The translational order parameter
disappears only in the two-phase region.

Because the translational order parameter $\psi_T$ is not equal to
zero in the hexatic phase, it is interesting to calculate the
coefficient of diffusion in order to be sure that the system is
really a liquid (but with quasi-long range orientational order).
It should be noted that in the cases when the hexatic phase was
reported (see, for example, \cite{foh1,foh2,foh5,prest1}) its
density range is extremely narrow, and it is very difficult to
calculate its dynamic properties. In the presence of quenched
disorder the hexatic phase considerably widens, and the
calculation of diffusion coefficient becomes more reliable. In
Fig.~\ref{fig:fig6} we present the diffusion coefficient for
$\sigma=1.15$ with and without random pinning. One can see that
without pinning the increase of the diffusion coefficient begins
at $\rho\approx 0.63$, while in the system with random pinning the
diffusion coefficient begins to become nonzero at $\rho\approx
0.65$ in accordance with the results shown in Figs. \ref{fig:fig4}
and \ref{fig:fig5}.

\subsection{$\sigma=1.35$}

In the case of $\sigma=1.35$, the results are more complex and
interesting in comparison with the previous Subsection. As it was
shown in our publications \cite{dfrt1,dfrt2,dfrt3,dfrt4,dfrt5},
the phase diagram of the system obtained with the help of the
Helmholtz free energy calculations for different phases and
construction of a common tangent to them \cite{book_fs} has three
different crystal phases, one of them has square symmetry and the
other two are the low density and high density triangular lattices
(see, for example, Fig. 6 (a) in \cite{dfrt3}). As it was
suggested in \cite{dfrt1,dfrt2,dfrt3,dfrt4}, melting of the low
density triangular phase is a continuous two-stage transition,
with a narrow intermediate hexatic phase. This conclusion was made
from the double-tangent construction and calculations of the
behavior of OOPs and TOPs. However, the more precise analysis
based on the study of the orientational and translational
correlation functions is necessary. At the same time at high
densities the square and triangular phases melt through the first
order transitions. At high temperatures and high density there is
one first order transition between triangular solid and isotropic
liquid. It should be noted that the water-like thermodynamic and
dynamic anomalies do exist in this case \cite{dfrt3}.

Let us first consider the effect of pinning on the high density
part of the phase diagram where melting occurs through the
first-order phase transition. Before to proceed, let us consider
the behavior of the system without quenched disorder in more
detail.

\begin{figure}%

\includegraphics[width=8cm]{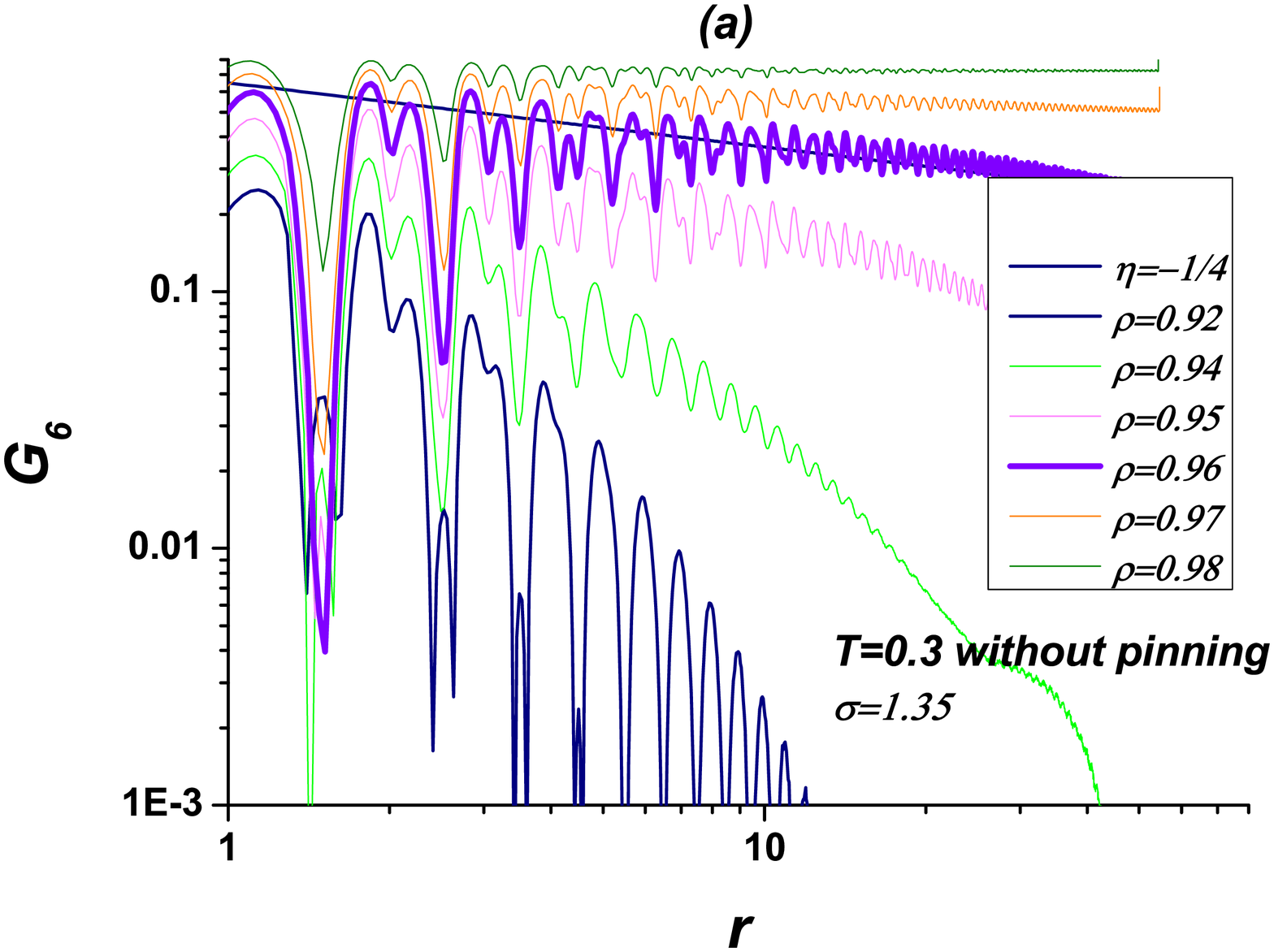}%

\includegraphics[width=8cm]{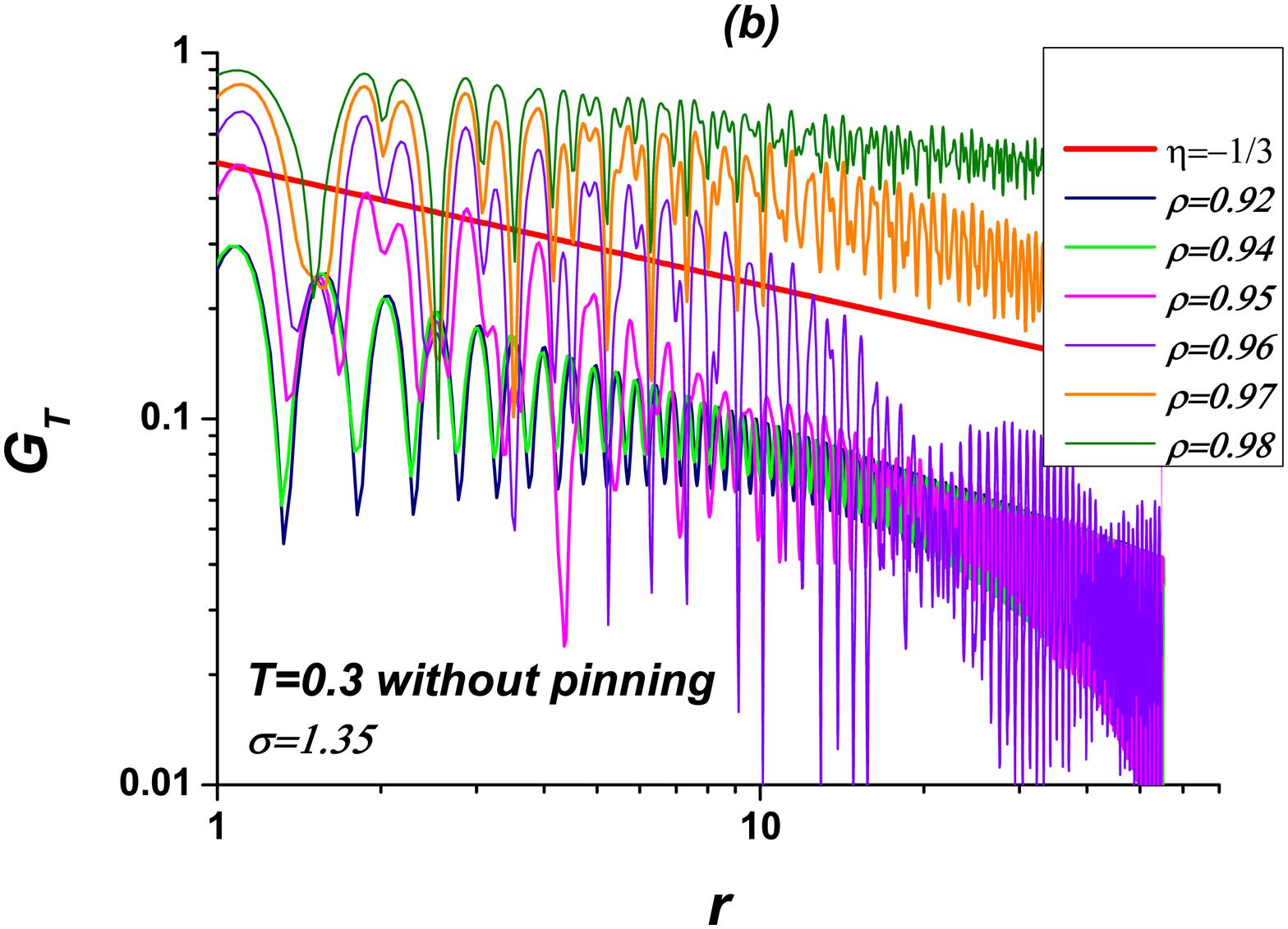}%

\includegraphics[width=8cm]{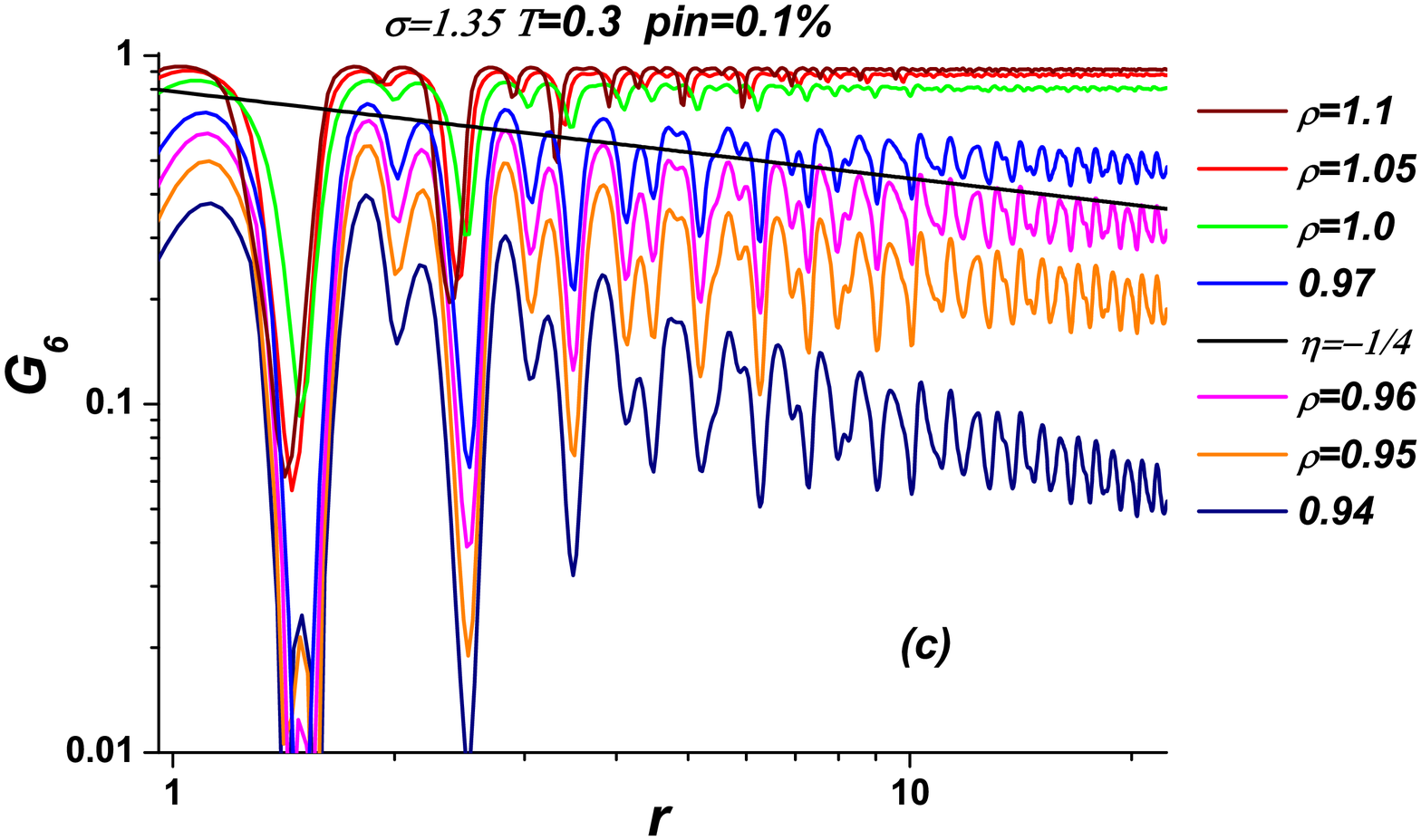}%

\includegraphics[width=8cm]{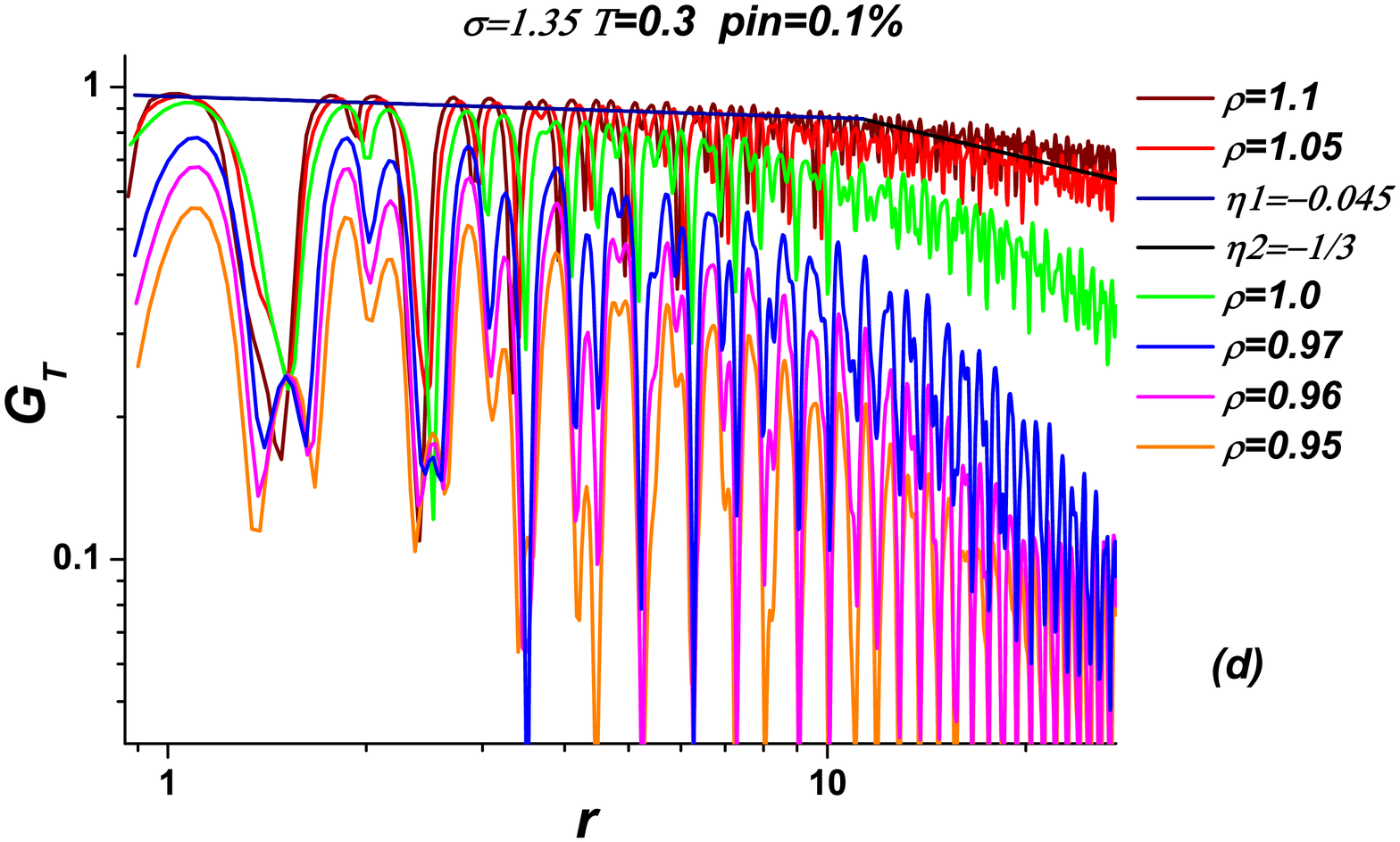}%

\caption{\label{fig:fig7}  OCF $G_6$ and TCF $G_T$ for different
densities at $T=0.3$ without random pinning ((a) and (b)) and
concentration of the pinning centers $0.1\%$ ((c) and (d)) for
$\sigma=1.35$.}
\end{figure}

\begin{figure}%

\includegraphics[width=8cm]{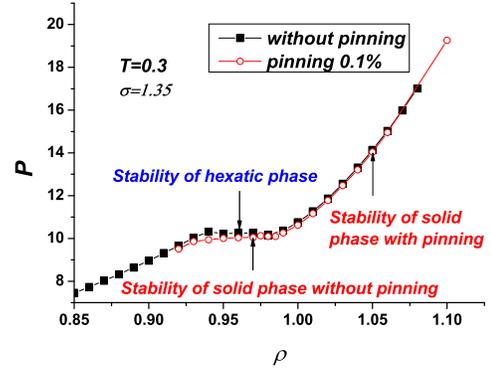}%

\caption{\label{fig:fig8}  Isotherms of the system with
$\sigma=1.35$ without pinning (squares) and with the concentration
of the pinning centers equal to $0.1\%$ (empty circles) for
$T=0.3$.}
\end{figure}

Fig.~\ref{fig:fig7} illustrates the behavior of OCF and TCF for
different densities at $T=0.3$ without pinning ((a) and (b)) and
with the concentration of the pinning centers equal to $0.1\%$
((c) and (d)). Using equations $\eta_T=1/3$ and $\eta_6=1/4$, we
calculated the limits of stability of hexatic and solid phases.
One can see that without quenched disorder both these limits lie
inside the two-phase region (Van der Waals loop) of the
corresponding equation of state (see Fig.~\ref{fig:fig8}). From
these results it is possible to conclude that in this case there
is the only first-order liquid-solid transition without hexatic
phase.

On the other hand, from Fig.~\ref{fig:fig7} (c) one can see that
random pinning leaves almost unaffected the limit of stability of
the hexatic phase, while considerably changes the location of
hexatic-solid transition (see Fig.~\ref{fig:fig7} (d)). In this
case the melting scenario qualitatively changes - instead of one
first-order transition we have two-stage melting with first-order
liquid-hexatic transition and continuous Kosterlitz-Thouless
hexatic-solid transition with the intermediate hexatic phase. For
the first time this kind of behavior was found in \cite{dfrt5}.

\begin{figure}%

\includegraphics[width=8cm]{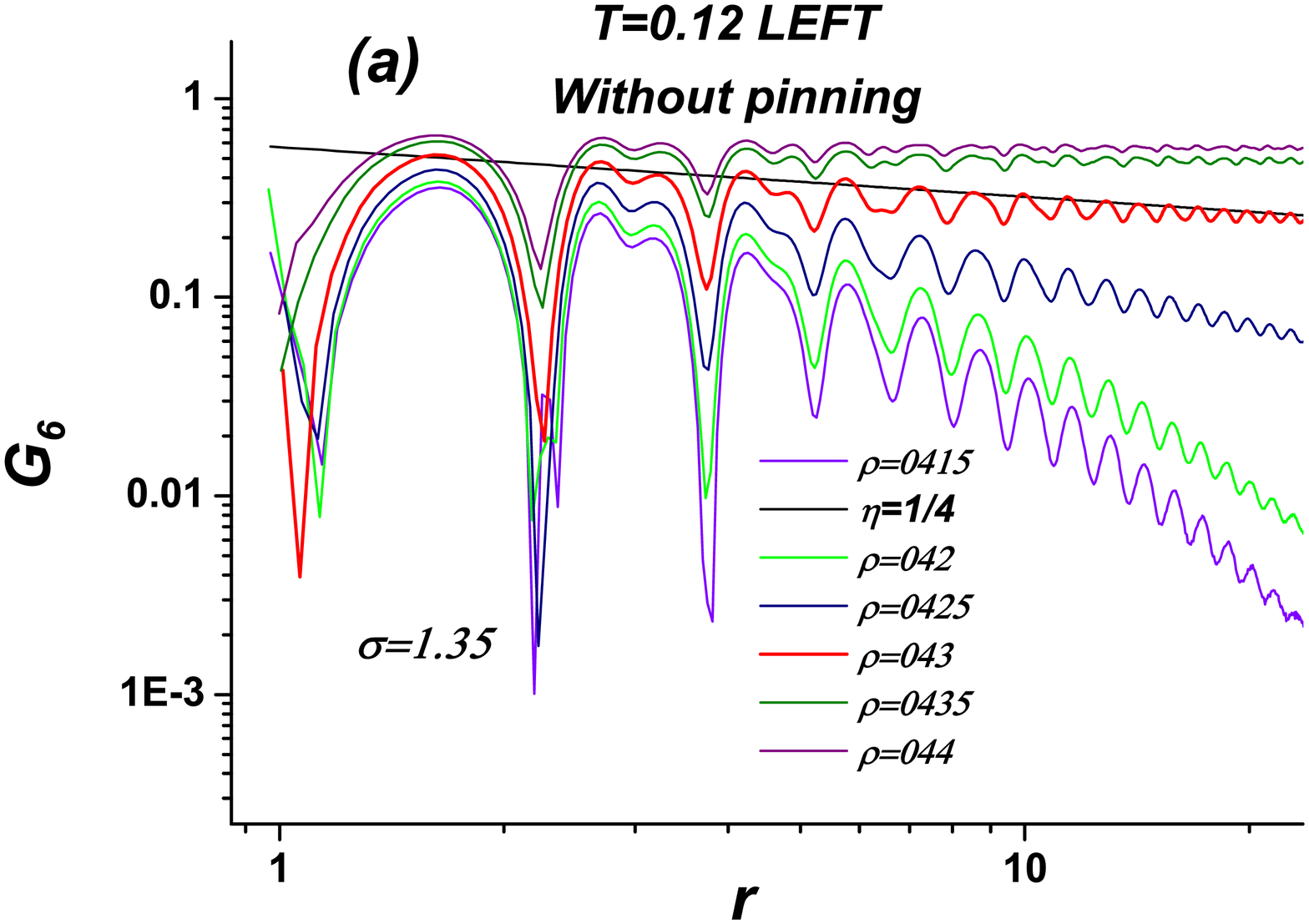}%

\includegraphics[width=8cm]{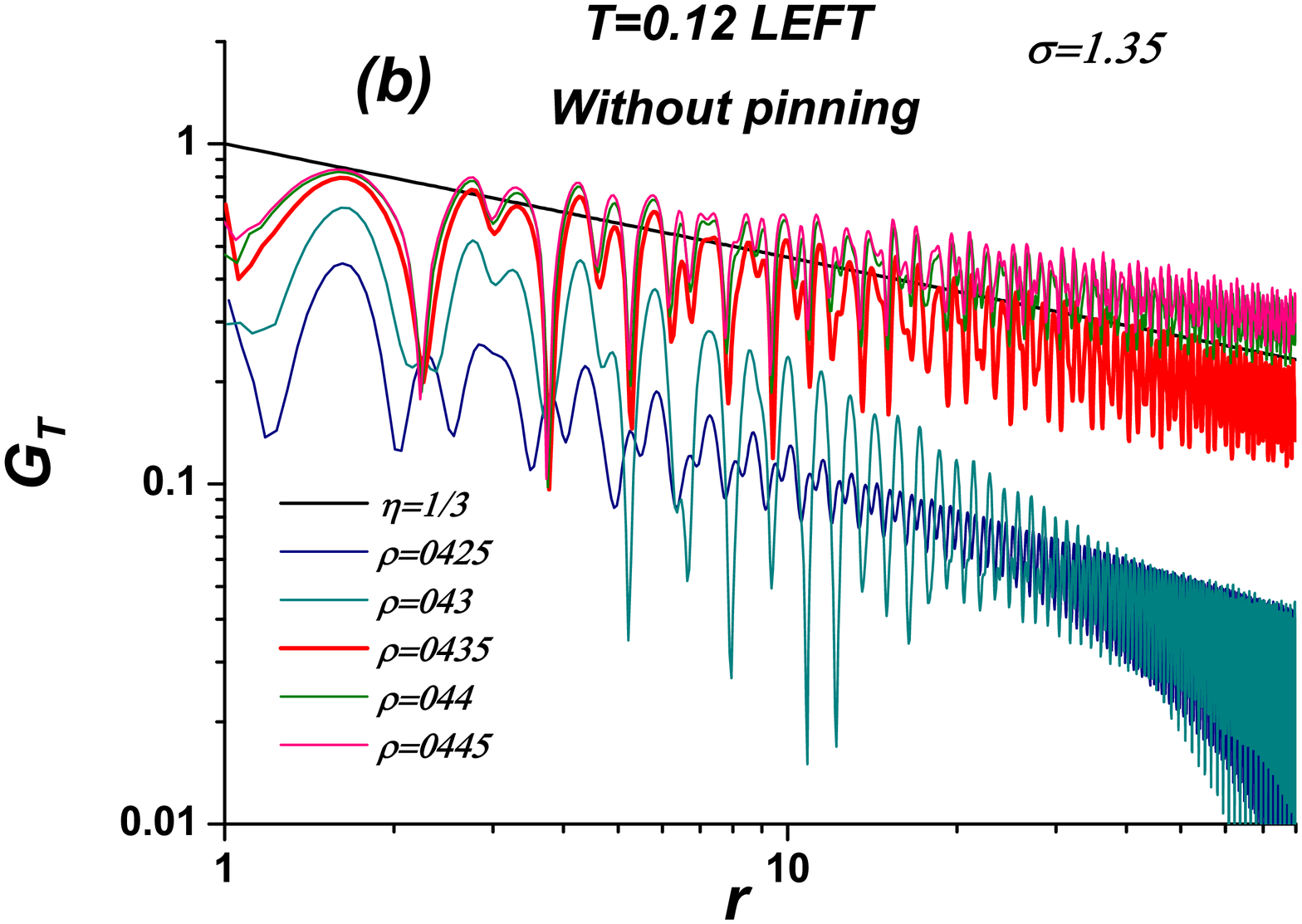}%

\includegraphics[width=8cm]{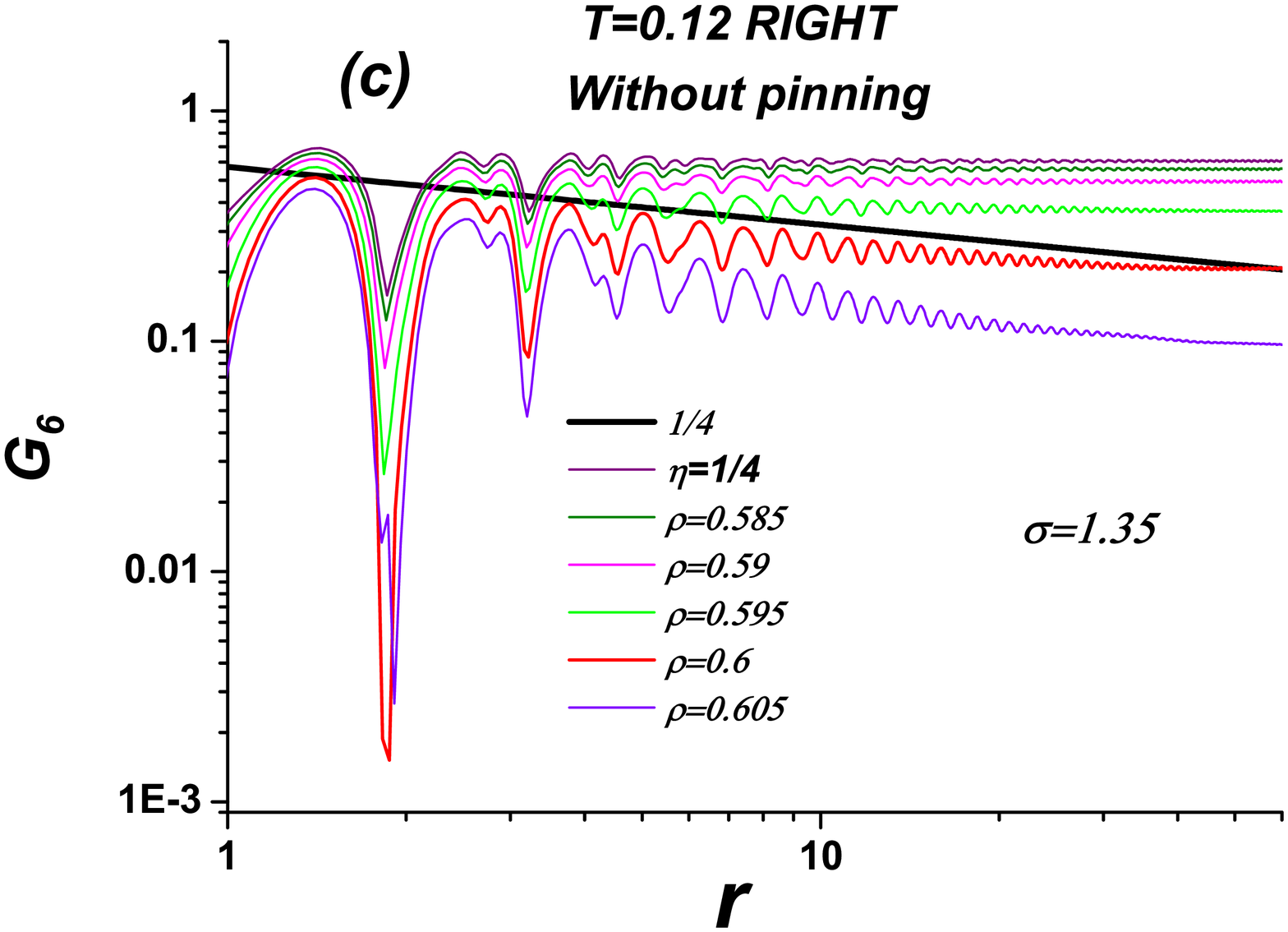}%

\includegraphics[width=8cm]{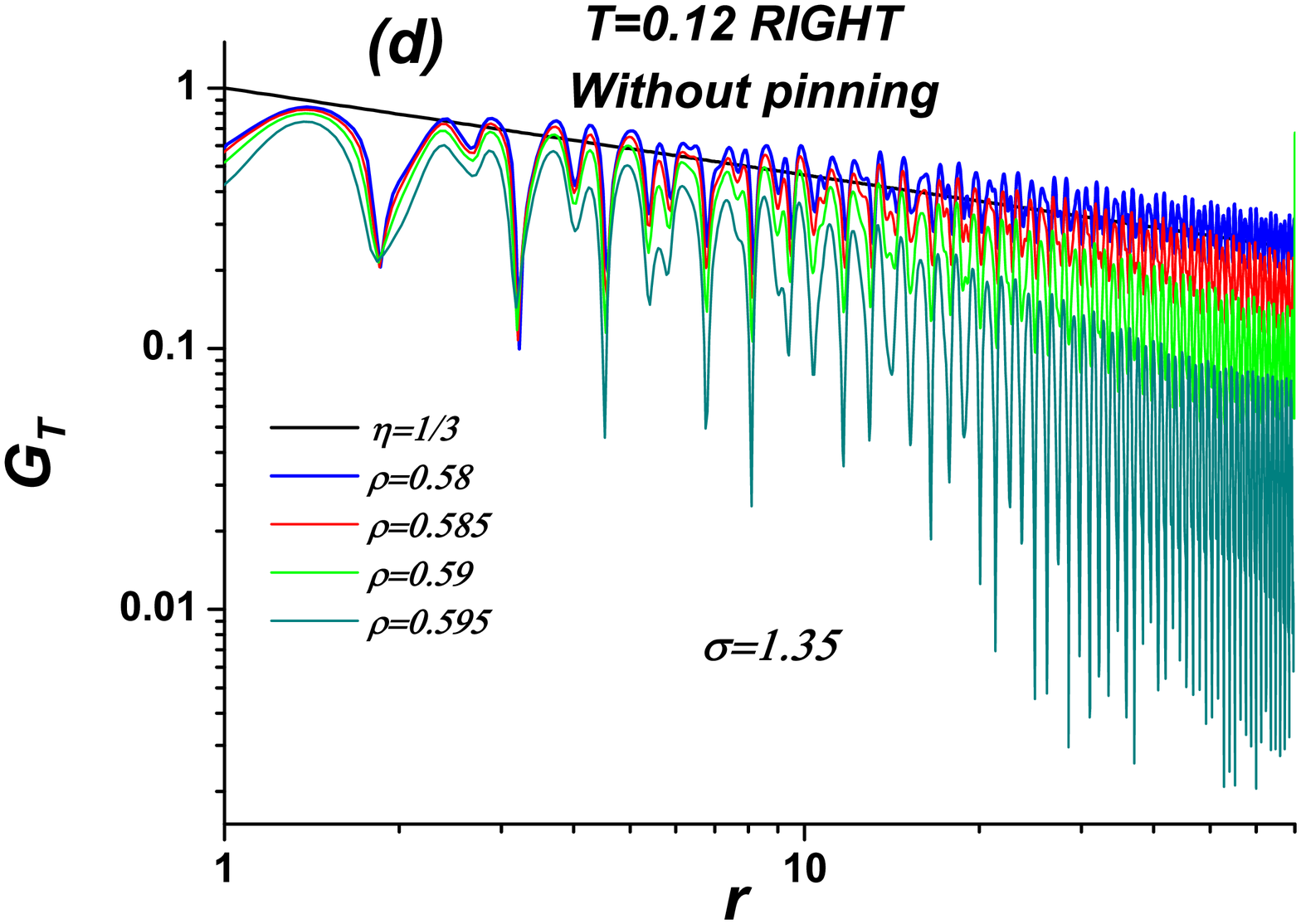}%

\caption{\label{fig:fig9}  OCF $G_6$ and TCF $G_T$ for different
densities at $T=0.12$ without random pinning for left ((a) and
(b)) and right ((c) and (d)) branches of dome at the low-density
part of the phase diagram  (see, for example, Fig. 6 (a) in
\cite{dfrt3}) for $\sigma=1.35$.}
\end{figure}

\begin{figure}%

\includegraphics[width=8cm]{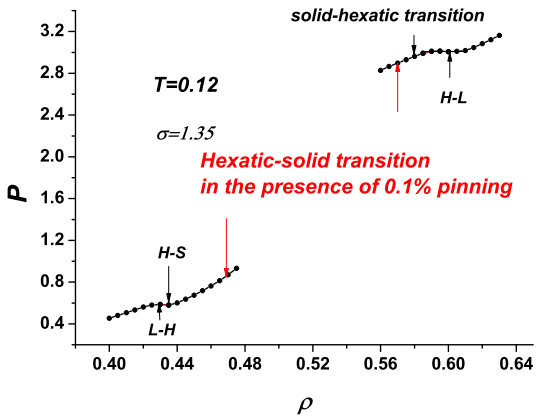}%

\caption{\label{fig:fig10} Isotherms of the system with
$\sigma=1.35$ without random pinning and with the concentration of
the pinning centers equal to $0.1\%$ for $T=0.12$.}
\end{figure}

Even more interesting picture takes place at low densities. In
Fig.~\ref{fig:fig9} we present the OCFs and TCFs for the system
without random pinning at $T=0.12$ at densities corresponding to
the left and right borders of the "cupola" of the triangular solid
at low density part of the phase diagram (see Fig. 6 (a) in
\cite{dfrt3} and Fig.~\ref{fig:fig12}). In Fig.~\ref{fig:fig10}
the isotherms corresponding to these borders are shown for the
same temperature $T=0.12$. One can see, that the limits of
stability determined from the equations $\eta_T=1/3$ for the solid
phase and $\eta_6=1/4$ for the hexatic phase are inside the
two-phase region of the first-order melting transition of the
system without pinning for the left part of the dome, while at the
right border of the dome the line of solid-hexatic transition
obtained from condition $\eta_T=1/3$ is outside of the two-phase
region (see Fig.~\ref{fig:fig10}). From this picture one can
conclude that without random pinning at the low density border of
the dome the system melts through one first-order transition,
while at high density (right) border the two-stage melting takes
place with first-order liquid-hexatic transition and continuous
hexatic-solid one.

As it was discussed above, random pinning leaves almost unaffected
the line of the liquid-hexatic transition, while considerably
changes the location of hexatic-solid stability limit. At the same
time, the Van der Waals loops which are characteristic of the
first order transition are almost unchanged by impurities. As a
result, in the presence of random pinning the hexatic phase
drastically widens in accordance with the theoretical predictions
\cite{nel_dis1,nel_dis2,keim3,keim4,dfrt5}, while in the case of
the single first-order melting the transitions splits into
first-order liquid-hexatic transition and continuous hexatic-solid
transition. The stability limits of the hexatic-solid transitions
at $T=0.12$ in the presence of the random pinning are shown in
Fig.~\ref{fig:fig10}.

\begin{figure}%

\includegraphics[width=8cm]{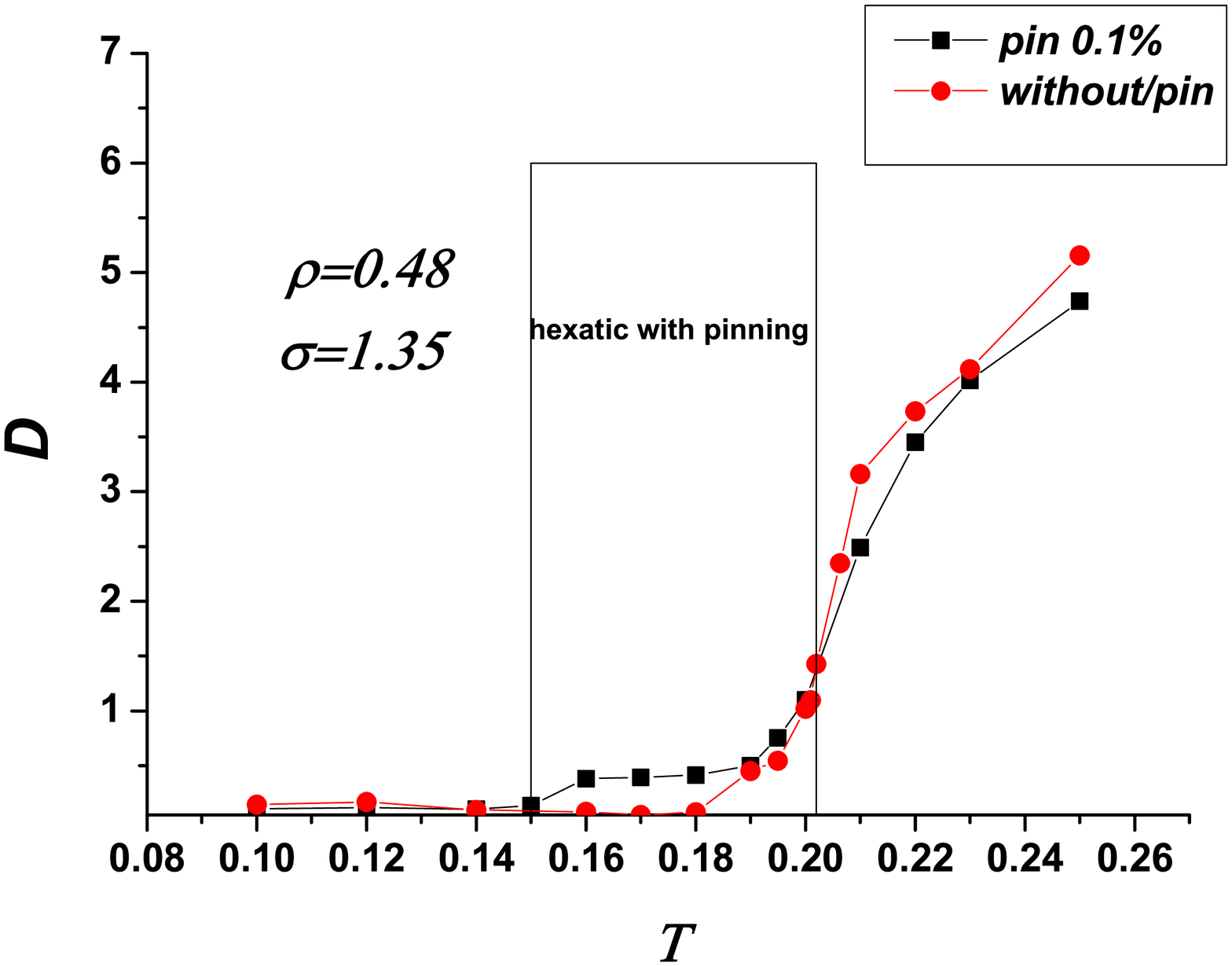}%

\caption{\label{fig:fig11} The diffusion coefficient of the system
with $\sigma = 1.35$ with (red circles) and without (black
squares) random pinning.}
\end{figure}

\begin{figure}%

\includegraphics[width=8cm]{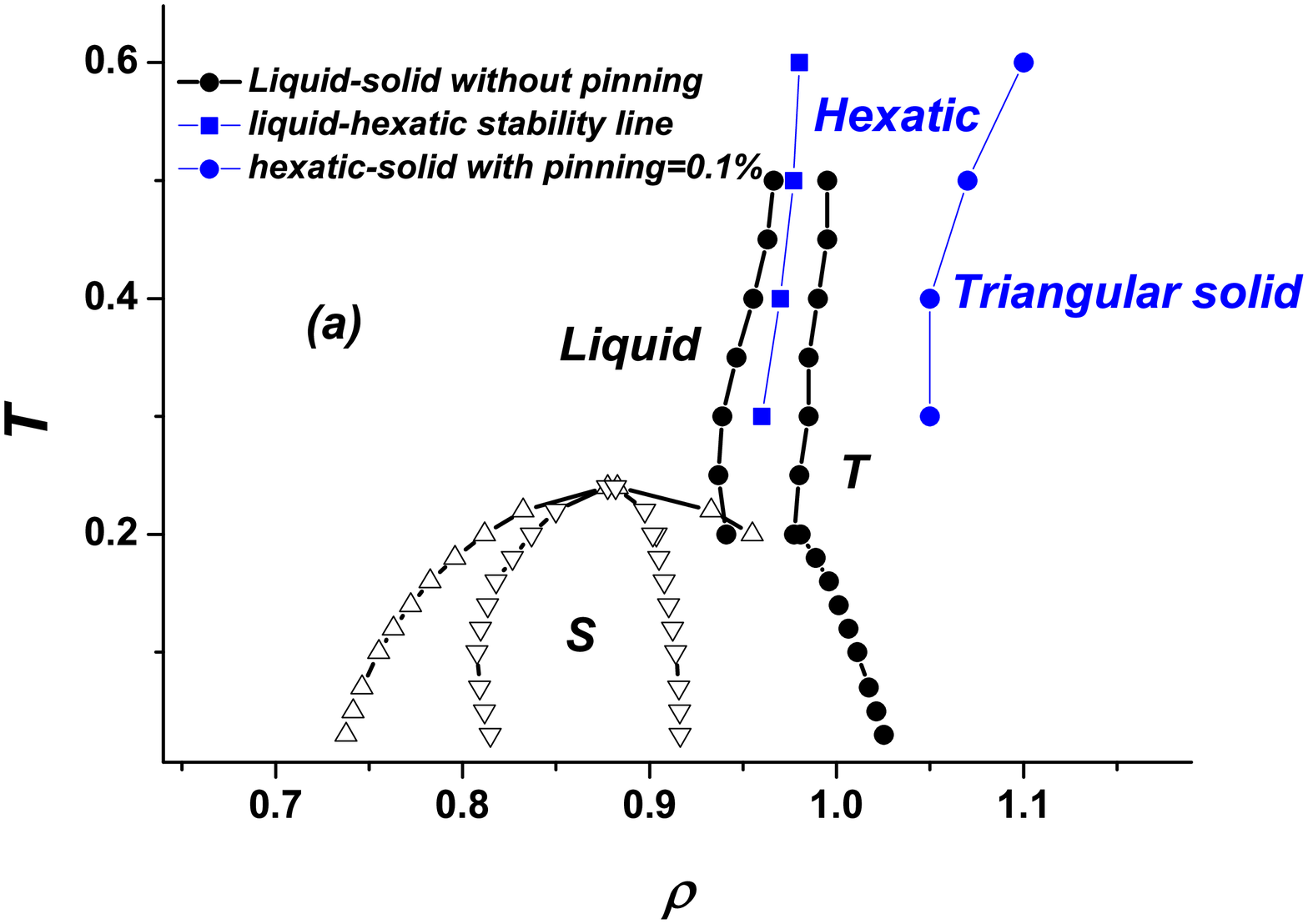}%

\includegraphics[width=8cm]{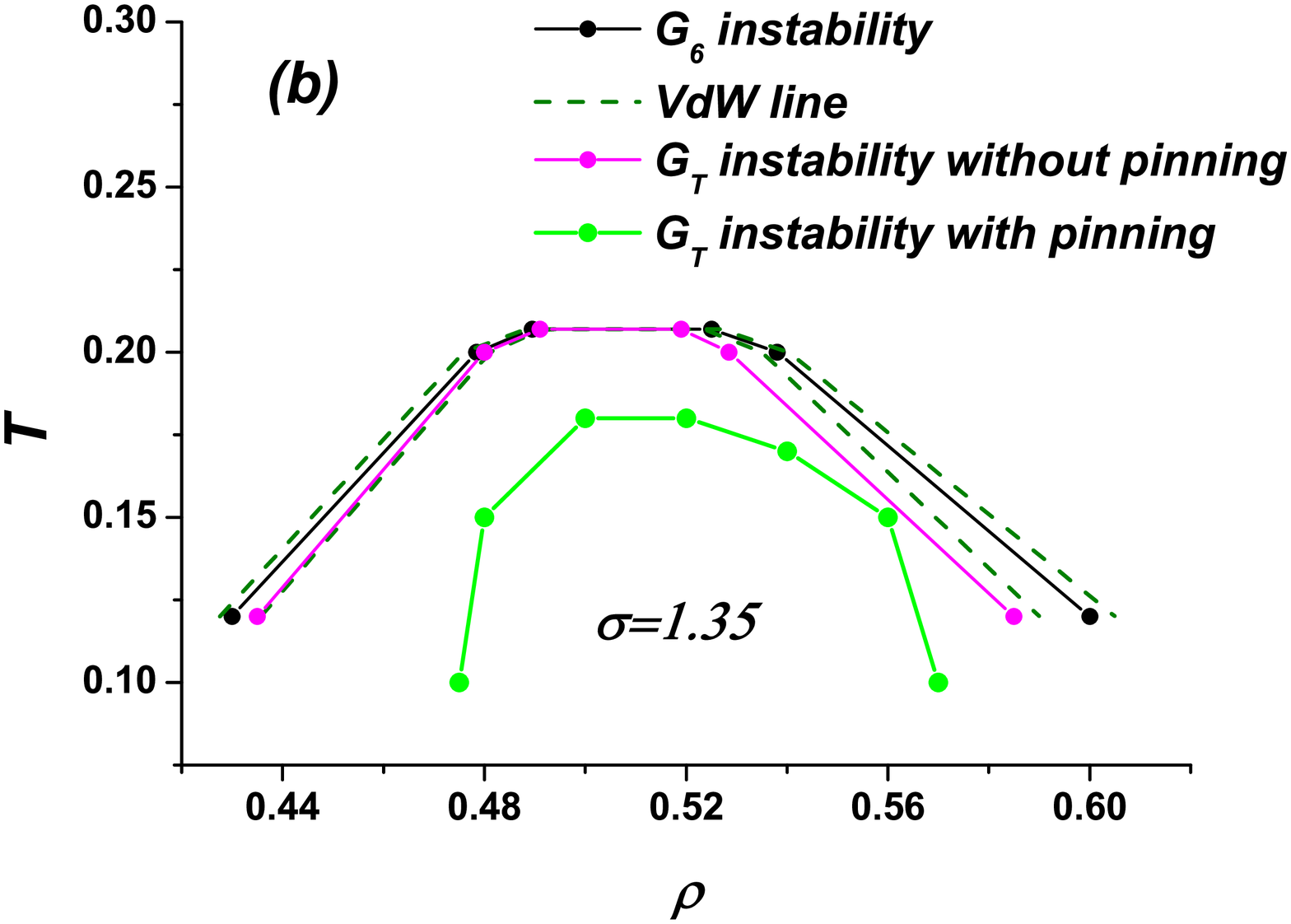}%

\caption{\label{fig:fig12} Phase diagram of the system with
$\sigma=1.35$: (a) high density part with first order solid-liquid
transition without random pinning (black lines) and two
transitions - first-order liquid-hexatic and continuous
hexatic-solid - for the concentration of the pinning centers equal
to $0.1\%$; (b) low-density part of the phase diagram. Dashed
lines corresponds to the first order transition calculated from
Maxwell construction for the Van der Waals loops (see
Fig.~\ref{fig:fig10}). Without random pinning melting occurs
through the first-order transition on the left branch of the dome,
while at right branch melting scenario corresponds to two
transitions - first-order liquid-hexatic and continuous
hexatic-solid. Green line demonstrates the shift of the
hexatic-solid transition in the presence of the random pinning.}
\end{figure}

As in the previous subsection, we calculated the diffusion
coefficient of the system without and with random pinning
(Fig.~\ref{fig:fig11}) for $\rho=0.48$. One can see that without
pinning the increase of the diffusion coefficient begins at
$T\approx 0.18$, while in the system with random pinning the
diffusion coefficient begins to become nonzero at $T\approx 0.15$
in accordance with the results shown in Fig.~\ref{fig:fig12} (b).

The results characterizing the influence of random pinning on the
phase diagram of the system with $\sigma=1.35$ are shown in
Fig.~\ref{fig:fig12}. The effect of random pinning on the
first-order melting transition is shown in Fig.~\ref{fig:fig12}
(a) for the temperatures higher than $0.3$. It should be noted
that the phase diagram was calculated using the double-tangent
construction to the Helmholtz free energy and the Maxwell
construction for the corresponding Van der Waals loops. In this
case, as it was discussed above, the transition in the pure system
is of first order. The line where $G_6(r)$ decays algebraically
with the exponent equal to $1/4$ is located inside the two-phase
region. At the same time, in the presence of random pinning the
line of solid-hexatic transition obtained from condition
$\eta_T=1/3$ is shifted to higher densities, and random pinning
transforms the single first-order transition into two transition
with rather wide hexatic phase. The solid-hexatic transition is
continuous, while the hexatic phase transforms into isotropic
liquid through first-order transition.

In Fig.~\ref{fig:fig12} (b) we plot the low-density part of the
phase diagram. The dashed lines correspond to the two-phase
regions obtained with the help of the Maxwell constructions for
the corresponding Van der Waals loops. One can see that without
random pinning the limits of stability of the hexatic phase
obtained from the equation $\eta_6=1/4$ are inside the two-phase
region for both the left and right borders of the dome, while the
hexatic-solid transition lines calculated from the equation
$\eta_T=1/3$ have different locations: it is inside the two-phase
region for the left border and outside for the right one. As it
was discussed above, in the left part of the phase diagram the
system melts through a single first-order transition, however, at
the left part melting occurs through two transitions with the
narrow intermediate hexatic phase and first-order liquid-hexatic
transition and continuous hexatic-solid one. In the presence of
quenched disorder the hexatic phase appears at the left part of
the phase diagram and considerably widens at the right part.

\section{Conclusions}

In conclusion, in this paper we present the computer simulation
study of melting transition of $2D$ core softened system with two
lengthes of the repulsive shoulder (Eq. (\ref{3}) and
Fig.~\ref{fig:fig1}) in the presence of random pinning. It is
shown that without the quenched disorder the system with small
repulsive shoulder $\sigma=1.15$ which is close in the shape to
the soft disks $1/r^n$ with $n=14$ melts in accordance with the
melting scenario proposed in Refs.~\cite{foh1,foh2,foh4}
(first-order liquid-hexatic and continuous hexatic-solid
transitions). Random pinning widens the hexatic phase, but leaves
the melting scenario unchanged.

Melting of the system with larger repulsive step $\sigma=1.35$ is
much more complex (see Fig.~\ref{fig:fig12}). At high densities
the conventional first-order transition takes place without random
pinning (Fig.~\ref{fig:fig12} (a)). However, disorder drastically
changes this melting scenario. The single first-order transition
transforms into two transitions, one of them (solid-hexatic) is
the continuous KTHNY-like transition, while the hexatic to
isotropic liquid transition occurs as the first order transition
in accordance with the recently proposed scenarios
\cite{foh1,foh2,foh4}. The possible mechanism for this transition
is the spontaneous proliferation of grain boundaries
\cite{chui83,foh3,foh5}. This scenario of melting in the presence
of pinning was proposed in Ref. \cite{dfrt5}. It should be noted,
that melting scenario with the single first-order transition
corresponds to the system at low enough temperature. At high
temperatures the repulsive shoulder of the potential becomes
ineffective, and the properties of the potential will be similar
to the soft disks $1/r^n$ with $n=14$. In this case one can expect
that there is critical temperature (some kind of tricritical
point) above which melting should occur through two transitions in
accordance with the scenario proposed in Ref. \cite{foh4}.

At low densities the system without pinning melts through the
single first-order transition at the left border of the dome
(Fig.~\ref{fig:fig12}) (b), while at the right border one can
observe the two-stage melting in accordance with the scenario
proposed in \cite{foh1,foh2,foh4}. The similar behavior was
recently found in the system of Herzian disks \cite{foh7}. The
random pinning transforms the first-order transition at left
border into two transitions with wide hexatic phase and widens the
hexatic phase at the right border.

It should be noted, that the nature of the first-order
liquid-hexatic transition is still puzzling, because the
conventional theory like the KTHNY one can not describe the
first-order liquid-hexatic transition.

The results of this study can be useful for the qualitative
understanding the behavior of water confined in the hydrophobic
slit nanopores \cite{nat1}, especially because the square solid
structure was found (see Fig.~\ref{fig:fig12}) that is similar to
the one discovered in experiment \cite{nat1}.

\bigskip

The authors are grateful to S.M. Stishov and V.V. Brazhkin for
valuable discussions. We thank the Russian Scientifc Center at
Kurchatov Institute and Joint Supercomputing Center of Russian
Academy of Science for computational facilities. The work was
supported by the Russian Science Foundation (Grant No
14-12-00820).

\end{document}